\documentclass[%
 reprint,
superscriptaddress,
 amsmath,amssymb,
  aps,
prb,
]{revtex4-1}
\pdfoutput=1 

\usepackage{graphicx}  
\usepackage{dcolumn}   
\usepackage{bm}        
\usepackage{verbatim}   
\usepackage{here} 
\usepackage[caption=false]{subfig}

\begin{document}

\title{An efficient method for calculating spatially extended electronic states\\ of large systems with a divide-and-conquer approach}

\author{Shunsuke Yamada}
\affiliation{Department of Physics, The University of Tokyo, Tokyo 113-0033, Japan}
\author{Fuyuki Shimojo}
\affiliation{Department of Physics, Kumamoto University, Kumamoto 860-8555, Japan}
\author{Ryosuke Akashi}
\affiliation{Department of Physics, The University of Tokyo, Tokyo 113-0033, Japan}
\author{Shinji Tsuneyuki}
\affiliation{Department of Physics, The University of Tokyo, Tokyo 113-0033, Japan}
\affiliation{Institute for Solid State Physics, The University of Tokyo, Kashiwa 277-8581, Japan}

\date{\today}
\begin{abstract}
We present an efficient post-processing method for calculating the electronic structure of nanosystems based on the divide-and-conquer approach to density functional theory (DC-DFT), 
in which a system is divided into subsystems whose electronic structure is solved separately.
In this post process, the Kohn-Sham Hamiltonian of the total system is easily derived from the orbitals and orbital energies of subsystems obtained by DC-DFT without time-consuming and redundant computation.
The resultant orbitals spatially extended over the total system are described as linear combinations of the orbitals of the subsystems.
The size of the Hamiltonian matrix can be much reduced from that for conventional calculation, so that our method is fast and applicable to general huge systems for investigating the nature of electronic states.
 \end{abstract}

\maketitle

\section{Introduction}

First principles calculations based on the density functional theory (DFT) \cite{hohenberg,kohn} have been widely used for investigating material properties and phenomena in condensed matter physics. While calculations with hundreds of atoms are currently routine, there is a demand for simulations of larger systems.
However, the computational time for conventional DFT calculations grows as O($N^3$) and such systems require massive computational effort which is often impossible to realize.
Linearly scaling methods, or O($N$) methods, for DFT allow us to study such large systems \cite{bowler}.
In particular, the divide-and-conquer approach to density functional theory (DC-DFT)
\cite{yang,dixon,shimojo2005,ozaki,kobayashi,shimojo2008,wang,zhao,ohba,shimojo2014}
has robust convergence properties.
In this approach, the total system is divided into overlapping subsystems often called fragments. 
The total energy is minimized through an iterative procedure of two steps: (i) Solving the Kohn-Sham equation for each fragment independently, and (ii) aggregating the electron densities of the fragments to calculate the electronic potential.

A difficult problem with the O($N$) methods is that the electronic structure calculation for the total system requires a time-consuming post-processing procedure. 
Generally, if we want to obtain the one-electron orbitals and orbital energies for the electronic state spatially extended over the total system,
it is necessary to diagonalize the Hamiltonian matrix for the total system with one-shot conventional O($N^3$) calculations by using the electron density obtained in the O($N$) procedure.
These calculations require a large number of basis functions and a time-consuming calculation for the Hamiltonian matrix elements.
For the former problem, a method is proposed to reduce the atomic-like basis functions to some extent \cite{nakata2014,nakata2015}.
The latter problem is due to that the matrix elements are calculated by explicitly treating the Hamiltonian operator.
In DC-DFT, this calculation seems redundant since the Kohn-Sham equations for overlapping fragments have already been solved in the DC-DFT calculation: 
The effects of kinetic and potential terms should in principle be embedded in the fragment orbitals.

The fragment molecular orbital (FMO) method \cite{kitaura} is one of the DC approaches specialized for covalent-bonding molecules.
There are some post-processing schemes of FMO to evaluate the electronic state of a whole molecule by utilizing information of the fragments \cite{inadomi, watanabe, fedorov, tsuneyuki, kobori}.
Among them, the FMO linear combination of molecular orbitals (LCMO) method \cite{tsuneyuki, kobori} is a particularly efficient scheme.
In FMO-LCMO, the one-electron Hamiltonian matrix is formulated and calculated using the molecular orbitals (MOs) of each fragment as basis functions, so that the wave function of the whole molecule is represented by a LCMO of the fragments without the redundant recalculation.
This scheme can reduce the dimension of the Hamiltonian matrix and the computational cost for calculating the matrix elements.
Nonetheless, this scheme cannot be adapted to general materials straightforwardly because FMO crucially relies on the specific property of the $sp^3$ orbitals of carbon atoms.

In this work, we present a new post-processing method of DC-DFT, which allows us to calculate the eigenstate for general materials utilizing output of the DC-DFT calculation.
To this end, we use DC-DFT with plane-wave basis functions, especially the lean divide-and-conquer (LDC) DFT \cite{shimojo2014}, that can be applied to general materials with a systematic procedure.
In our method, a small number of basis functions are constructed by reducing the fragment orbitals derived from DC-DFT.
The total Hamiltonian matrix defined by the localized basis set is derived easily from the fragment orbitals and their orbital energies in the manner of FMO-LCMO.
Each wave function of the whole system is represented by a linear combination of the fragment orbitals (LCFO).
The diagonalization process can be done with little computational cost thanks to the reduced number of the basis functions.
Thus our procedure, called DC-LCFO hereafter, has advantages of a low cost and versatility.

This paper is organized as follows. Section \ref{sec:formalism} describes DC-DFT and our DC-LCFO.
Section \ref{sec:results} presents the computational results.
We analyze the parameter dependence for resultant eigenenergies and the feasibility of the matrix-size reduction of the Hamiltonian in our scheme.
We also compare the eigenenergies and the wave functions with those of the conventional method in P-doped Si and InGaN/GaN superlattice systems.
Finally, conclusion is given in Sec. \ref{sec:conclusion}.
In appendix \ref{appendix}, we compare the formulation of DC-LCFO with that of FMO-LCMO.

\section{Formalism \label{sec:formalism}}

\subsection{Divide-and-conquer density functional theory}

\begin{figure}
	\includegraphics[keepaspectratio, scale=0.8]{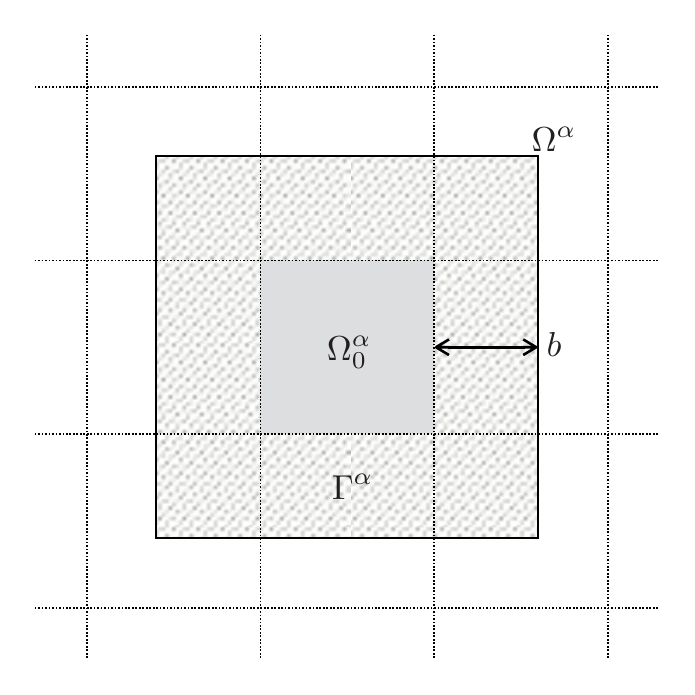}
	\caption{\label{fig:dc} 
		Schematic 2D image of the DC algorithm. 
		The whole system $\Omega$ is represented as a union of non-overlapping core domains 
		$\{ \Omega_{0}^{\alpha} \}$.
		Physical quantities of $\Omega_{0}^{\alpha}$ are evaluated on the fragment 
		$\Omega^{\alpha} = \Omega_{0}^{{\alpha}} \bigcup \Gamma^{\alpha}$
		, where  $\Gamma^{\alpha}$ is a buffer layer whose thickness is b.
	}
\end {figure}

We first review the fundamental formalism of DC-DFT \cite{yang,dixon,shimojo2005,ozaki,kobayashi,shimojo2008,wang,zhao,ohba,shimojo2014}.
In DC-DFT, the physical space $\Omega$ is represented as a union of non-overlapping core domains, $\Omega=\bigcup_{\alpha} \Omega_{0}^{\alpha}$, where $\Omega_{0}^{\alpha} \bigcap \Omega_{0}^{\beta}=\emptyset \, ({\alpha} \neq \beta) $. 
An additive quantity defined for $\Omega$ is described as the sum of those for the respective domains $\Omega_0^{\alpha}$.  
Practically, the summands are approximately evaluated by constructing modified domains, or fragments $\Omega^{\alpha} = \Omega_{0}^{{\alpha}} \bigcup \Gamma^{\alpha}$.
Here, $\Gamma^{\alpha}$ is the buffer region surrounding $\Omega_{0}^{\alpha}$ (Fig.~\ref{fig:dc}). Thus the electron density of the whole system $\Omega$ is given by
\begin{equation}
	\rho({\bf r}) = \sum_{\alpha} \bar{\rho}^{\alpha} ({\bf r}) ,
	\quad {\rho}^{\alpha} ({\bf r})= \sum_i f(\varepsilon^{\alpha}_i-\mu)| \phi^{\alpha}_i({\bf r})|^2 , 
	\label{eq:rho_tot}
\end{equation}
where $\bar{\rho}^{\alpha} ({\bf r})$ is the electron density in $\Omega^\alpha_0$ clipped from that of the $\alpha$th fragment $\rho^\alpha(r)$: $ \bar{\rho}^{\alpha} ({\bf r})={\rho}^{\alpha} ({\bf r})$ for ${\bf r} \in \Omega_{0}^{\alpha}$, while $=0$ otherwise.
The orbitals $\{\phi^{\alpha}_i({\bf r}), \, {\bf r} \in \Omega^{\alpha}\}$ and corresponding eigenenergies $\{\varepsilon^{\alpha}_i\}$ for each fragment are evaluated from the fragment Kohn-Sham (KS) equation,
\begin{equation}
	\hat{H}^{\alpha} |\phi^{\alpha}_i\rangle \equiv \left[ - \frac{1}{2} \nabla^2 + \hat{V}_{\mathrm{KS}}
	+  \hat{v}^{\alpha}_{\mathrm{bc}} \right] |\phi^{\alpha}_i\rangle
	 =\varepsilon^{\alpha}_i |\phi^{\alpha}_i\rangle ,
	 \label{eq:fragmentKSeq}
\end{equation}
with the boundary potential $v^{\alpha}_{\mathrm{bc}}({\bf r})$ that represents artificial effects of the buffer regions such as termination of the bonds and the boundary condition at $\partial \Omega^{\alpha}$.
Here $\hat{V}_{\mathrm{KS}}=\hat{V}_{\mathrm{KS}}[\rho]$ is the KS potential with the density of the total system.
The Fermi energy, or the chemical potential, $\mu$ in the Fermi distribution function $f$ is determined by the electron number condition $N=\int d^3 r \rho({\bf r}) $ imposed on the whole system. 
Here, the chemical potential $\mu$ in Eq.~\eqref{eq:rho_tot} is common for all the fragments and it is solved by the bisection or Newton-Raphson method \cite{shimojo2005, shimojo2008}.

\subsection{Strategy \label{subsec:strategy}}

Here we briefly describe the strategy and basic ideas in DC-LCFO. 
There are several methods for DC-DFT with different basis functions and different ways of handling the boundary effects of the fragments: their accuracy depends on the target systems.
In the present paper, we adopt LDC-DFT \cite{shimojo2014} as the basis, which utilizes the density template potential \cite{ohba} as $v^{\alpha}_{\mathrm{bc}}({\bf r})$ and the periodic boundary conditions at $\partial \Omega^{\alpha}$ on the fragment KS orbitals.
This method is suitable for calculations of condensed matter since it has relatively small overhead and convenient buffer configurations for calculations (see Sec.~\ref{subsec:hamiltonian}).
Note that we use the step-wise projection in Eq.~\eqref{eq:rho_tot} for simplicity \cite{ohba}, though continuous weight functions are used in the original LDC-DFT paper. 

Our fundamental assumption is that the eigenstates near the Fermi level can be well represented by patching a small number of fragment orbitals in the corresponding energy region \cite{tsuneyuki,kobori}. 
Thereby we shall first develop an algorithm to generate basis functions from the fragment orbitals $\{\phi^{\alpha}_i\}$ within the low-energy region.

As the next step, we shall introduce a method to construct the total Hamiltonian matrix using the basis functions and the fragment Hamiltonians $\{ \hat{H}^{\alpha} \}$ defined in Eq.~\eqref{eq:fragmentKSeq}.
The matrix elements of the Hamiltonian can be constructed simply by inner products among the fragment orbitals.
Hence the Hamiltonian matrix can be obtained without time-consuming calculations even when the exact exchange potential is taken into account.

The resultant basis functions are defined on each core domain $\Omega_{0}^{\alpha}$ and therefore it is not necessary to consider the total overlap matrix of the whole system.
Moreover, the Hamiltonian matrix has a far smaller dimension than the plane-wave basis case because the new basis functions are made of the fragment orbitals in the low-energy region.
Notably, a typical number of the basis functions per atom for practical accuracy is roughly 10--20, which is comparable to the atomic-like basis case \cite{bowler}. 
However the latter case has drawbacks such as the lack of systematic convergence.

There is a similar approach that utilizes the KS orbitals of subsystems as a basis set for evaluating the Green's functions, though the method is specialized for a quasi-1D system \cite{varga}.
In contrast, our scheme is the post-processing method of DC-DFT for a direct diagonalization of the Hamiltonian matrix.
DC-DFT can be systematically applied to general 3D systems.

\subsection{Basis set \label{subsec:basis}}

The basis functions of the present method are constructed as follows.
We introduce a cutoff energy $\varepsilon_{\mathrm{cut}}$ for the energy eigenvalues of the fragment orbitals in order to  restrict the number of the fragment orbitals used for constructing the basis set:
\begin{equation}
	\phi^{\alpha}_i({\bf r}),\, (i=1,\cdots,N_{\alpha}) ,
\end{equation}
where $N_{\alpha}$ is the number of the fragment orbitals satisfying $\varepsilon^{\alpha}_i< \varepsilon_{\mathrm{cut}}$, and $i$ is the orbital index.

 In order to eliminate a redundant contribution from the buffer region, we project the fragment wave functions onto $ \Omega_{0}^{\alpha}$, 
\begin{equation}
	|\phi^{\alpha}_i\rangle \longrightarrow |\bar{\phi}^{\alpha}_i\rangle=
	\int_{{\bf r} \in \Omega_{0}^{\alpha}} d^3 r 
	|{\bf r}\rangle\langle{\bf r}|\phi^{\alpha}_i\rangle. 
\end{equation}
To avoid overcompleteness with the projected orbitals, we construct a smaller set of orbitals from them. Namely, we define an overlap matrix within each fragment $\alpha$,
\begin{equation}
	S^{\alpha}_{ij} = \langle \bar{\phi}^{\alpha}_i | \bar{\phi}^{\alpha}_j \rangle
	 ,\quad (i,j=1,\cdots,N_{\alpha}).
\end{equation}
Next, we diagonalize it,
\begin{equation}
	S^{\alpha} \longrightarrow (U^{\alpha})^{\dagger} S^{\alpha} U^{\alpha} = 
	\mathrm{diag}(\lambda^{\alpha}_1,\lambda^{\alpha}_2,
	\cdots,\lambda^{\alpha}_{M_{\alpha}},0,0,\cdots),
	\label{eq:diag_s}
\end{equation}
where $U^{\alpha}$ and $\lambda^{\alpha}_i$ are the transformation matrix and the eigenvalue of $S^{\alpha}$, respectively, and $M_{\alpha} \equiv \mathrm{rank}\,S^{\alpha}$ is the number of the linearly independent eigenvectors.
Practically, we set a sufficiently small cutoff parameter $\lambda_{\mathrm{cut}}$ for the eigenvalues $\lambda^{\alpha}_i(>\lambda_{\mathrm{cut}})$ in order to control $M_{\alpha}$ .

The new basis functions are defined as,
\begin{equation}
	|\lambda^{\alpha}_i \rangle = \frac{1}{\sqrt{\lambda^{\alpha}_i}}
	\sum_{j=1}^{N_{\alpha}} | \bar{\phi}^{\alpha}_j \rangle U^{\alpha}_{ji}, \quad (i=1,\cdots,M_{\alpha}).
	\label{eq:basis}
\end{equation}
It is notable that these basis functions are orthonormal:
\begin{equation}
	\langle \lambda^{\alpha}_i | \lambda^{\beta}_{j} \rangle =\delta_{{\alpha},\beta} \delta_{i,j} .
\end{equation}

\subsection{Hamiltonian matrix \label{subsec:hamiltonian}}

\begin{figure*}
	\subfloat[]{
		\label{fig:truncation}
		\includegraphics[keepaspectratio, scale=0.8]{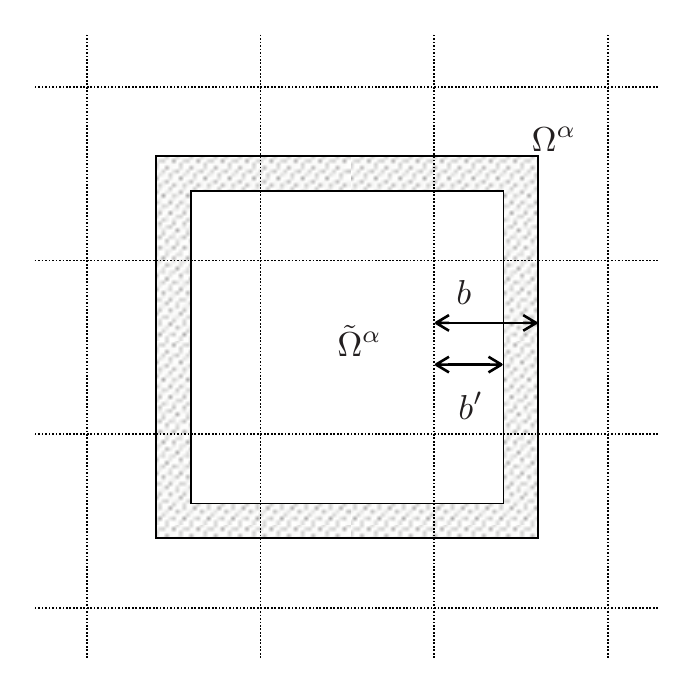}
	}
	\subfloat[]{
		\label{fig:beta}
		\includegraphics[keepaspectratio, scale=0.8]{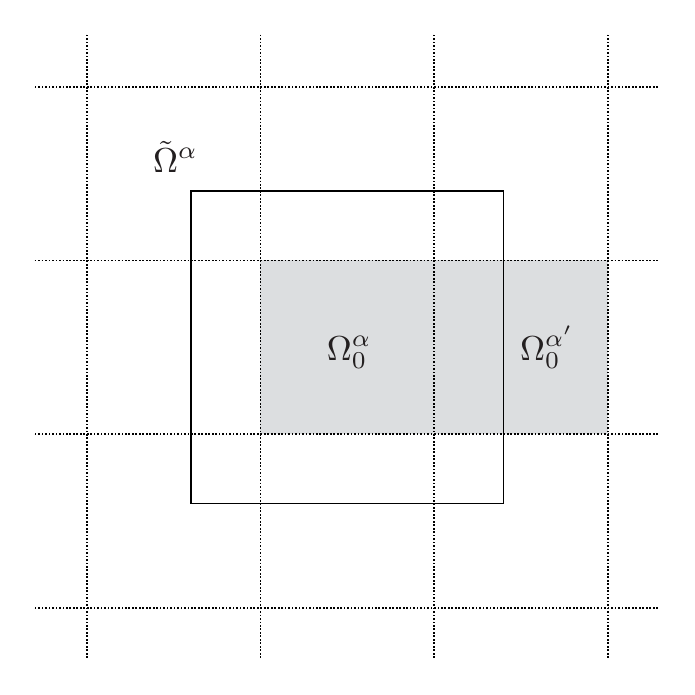}
	}
	\subfloat[]{
		\label{fig:beta-other}
		\includegraphics[keepaspectratio, scale=0.8]{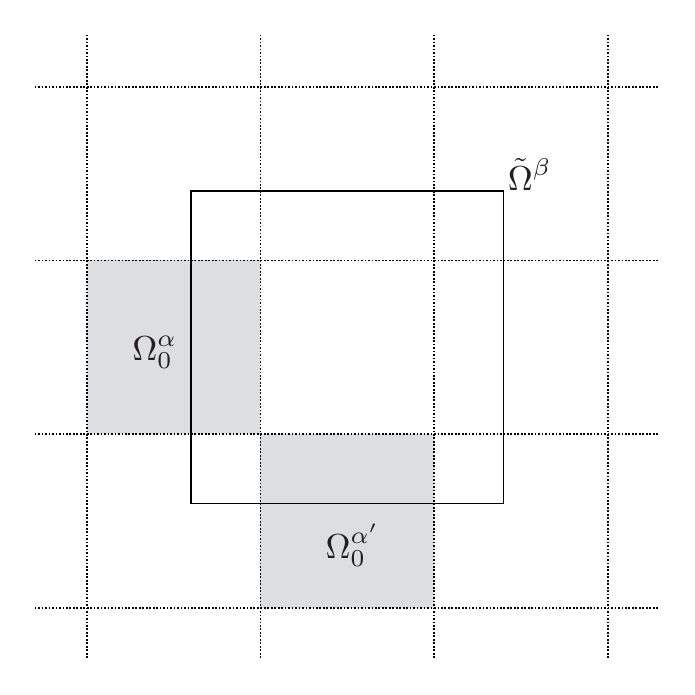}
	}
	\caption{
		2D schematic of the Hamiltonian matrix construction.
		(a) To truncate the Hamiltonian, we introduce the projection operator $\hat{P}^{\alpha}$
		 corresponding to $\tilde{\Omega}^{\alpha}$. 
		The buffer thickness $b'$ of $\tilde{\Omega}^{\alpha}$ is restricted in $0 \leq b' \leq b$.
		(b) The projection operator $\hat{P}^{\alpha}$ is inserted in
		$\langle \lambda^{\alpha'}_{i'} | \hat{H}| \lambda^{{\alpha}}_{i}\rangle$,
		where $\alpha$ and $\alpha'$ are the first-nearest-neighbor fragments.
		(c) The projection operator $\hat{P}^{\beta}$ is inserted in
		$\langle \lambda^{\alpha'}_{i'} | \hat{H}| \lambda^{{\alpha}}_{i}\rangle$,
		where $\alpha$ and $\alpha'$ are the second-nearest-neighbor fragments in the 2D system,
		and $\beta$ is the fragment overlapping with both ${\alpha}$ and ${\alpha}'$ (see Appendix).
	}
\end{figure*}

We shall construct the Hamiltonian matrix
$H_{\alpha' i', \alpha i} \equiv \langle \lambda^{\alpha'}_{i'} | \hat{H}| \lambda^{{\alpha}}_{i}\rangle$, where $\hat{H}=- \frac{1}{2} \nabla^2+\hat{V}_{\mathrm{KS}}$ 
is the conventional KS Hamiltonian operator of the total system. 
A straightforward approach to this calculation is real-space integeration, for example, $\langle \lambda^{\alpha'}_{i'} | \hat{V}^{\mathrm{local}}_{\mathrm{KS}}| \lambda^{{\alpha}}_{i}\rangle=\int d^3 r \langle \lambda^{\alpha'}_{i'} |{\bf r}\rangle {V}^{\mathrm{local}}_{\mathrm{KS}} ({\bf r}) \langle {\bf r}| \lambda^{{\alpha}}_{i}\rangle$, etc.
In the below, however, we show the Hamiltonian matrix elements can be evaluated from the fragment orbital energies without the numerically demanding integration.

The local part of the Hamiltonian has only diagonal blocks (${\alpha}'={\alpha}$) because of the non-overlapping nature of $|\lambda^{\alpha}_i\rangle$.
On the other hand, the nonzero off-diagonal blocks (${\alpha}'\neq{\alpha}$) come from the kinetic term $- \frac{1}{2} \nabla^2$ and the non-local potential term. 
In the DC scheme, it is assumed that the non-local terms decay within 
the buffer region $\Gamma^\alpha$. It means that the terms decay within 
neighbor core domains, because the thickness of $\Gamma^{\alpha}$ is assumed as the same or the half length of each core domain $\Omega^{\alpha}_{0}$ \cite{prodan-kohn, shimojo2014}.
Following this assumption, we consider only the diagonal blocks ($\alpha'=\alpha$) and the off-diagonal blocks ($\alpha' \neq \alpha$) between the face-, edge- and corner-sharing neighboring core domains.

We shall obtain an approximate expression for the Hamiltonian matrix, which is represented by the KS orbitals and eigenenergies of a certain fragment overlapping with the core domains $\Omega_0^{\alpha}$ and $\Omega_0^{\alpha'}$. 
The apparent absence of the Hamiltonian operator in the resulting expression allows us a consistent treatment regardless of whether or not whether numerically demanding non-local operators (e.g., exact exchange operator) are considered in the calculation within each fragment. 
The pivotal approximation is that the basis function $|\lambda^{\alpha}_{i}\rangle$ does not spill far out of the core domain through the Hamiltonian operation. 

Now we define a projection operator for the Hamiltonian decomposition,
\begin{equation}
	\hat{P}^{\alpha} = \int_{{\bf r} \in \tilde{\Omega}^{\alpha}} d^3 r  |{\bf r}\rangle\langle{\bf r}|,
\end{equation}
where $\tilde{\Omega}^{\alpha}$ is a region to truncate the spillage of the basis functions upon the Hamiltonian operation (Fig.~\ref{fig:truncation}).
Using the projection operator, we get the following exact expression,
\begin{eqnarray}
	\hat{H}|\lambda^{\alpha}_i\rangle 
	&=& \hat{P}^{\alpha} \hat{H}|\lambda^{\alpha}_i\rangle 
	+\hat{Q}^{\alpha} \hat{H}|\lambda^{\alpha}_i\rangle
	\nonumber \\
	&=& \hat{P}^{\alpha} \hat{H} \hat{P}^{\alpha}| \lambda^{\alpha}_{i} \rangle 
	+ \hat{Q}^{\alpha} \hat{H} | \lambda^{\alpha}_{i} \rangle,
	\label{eq:pq}
\end{eqnarray}
where $\hat{Q}^{\alpha}=\hat{1}-\hat{P}^{\alpha}$.
Furthermore, the projected total Hamiltonian $ \hat{P}^{\alpha} \hat{H} \hat{P}^{\alpha}$ can be converted to the fragment Hamiltonian $\hat{P}^{\alpha}\hat{H}^{\alpha} \hat{P}^{\alpha}$ due to a feature of $v^{\alpha}_{\mathrm{bc}}({\bf r})$ that vanishes at the core domain 
$\Omega^{\alpha}_0$.
The second term is presumably small if the range of $\tilde{\Omega}^{\alpha}$ is sufficiently larger than ${\Omega}^{\alpha}_{0}$.
Therefore the Hamiltonian $\hat{H}$ acting on the vector $|\lambda^{\alpha}_i\rangle$ can be reasonably approximated as $\hat{P}^{\alpha} \hat{H}^{\alpha} \hat{P}^{\alpha}$.

Thus we get to the following approximate form of the Hamiltonian matrix element,
\begin{equation}
	H_{\alpha' i', \alpha i} \approx
	\langle \lambda^{\alpha'}_{i'} | \hat{P}^{\alpha} \hat{H}^{\alpha} \hat{P}^{\alpha} | 
	\lambda^{{\alpha}}_{i} \rangle.
	\label{eq:hamiltonian}
\end{equation}
From the range of the projector $\hat{P}^{\alpha}$, it is obviously nonzero only for $\alpha'$ whose core domain ${\Omega}_{0}^{\alpha'}$ is overlapping with $\tilde{\Omega}^{\alpha}$ (Fig.~\ref{fig:beta}).
Since the fragment Hamiltonian $\hat{H}^{\alpha}$ can be represented through the fragment orbitals $\{\phi^{\alpha}_i\}_{i=1}^{\infty}$, we obtain,
\begin{eqnarray}
	\langle \lambda^{\alpha'}_{i'} | \hat{P}^{\alpha} \hat{H}^{\alpha} \hat{P}^{\alpha}
	 | \lambda^{{\alpha}}_{i} \rangle 
	&=& \langle \lambda^{\alpha'}_{i'} | \hat{P}^{\alpha} \left( 
	\sum_{j=1}^{\infty} \varepsilon^{\alpha}_j  | {\phi}^{\alpha}_j \rangle \langle {\phi}^{\alpha}_j |
	\right) \hat{P}^{\alpha}| \lambda^{{\alpha}}_{i} \rangle \nonumber \\
	&=& \sum_{j=1}^{\infty} \varepsilon^{\alpha}_j \langle \lambda^{\alpha'}_{i'} 
	 | \tilde{\phi}^{\alpha}_j \rangle \langle 	
	\tilde{\phi}^{\alpha}_j | \lambda^{{\alpha}}_{i} \rangle \nonumber \\
	&\approx& \sum_{j=1}^{N_{\alpha}} \varepsilon^{\alpha}_j \langle \lambda^{\alpha'}_{i'} 
	 | \tilde{\phi}^{\alpha}_j \rangle \langle 	
	 \tilde{\phi}^{\alpha}_j | \lambda^{{\alpha}}_{i} \rangle,
	 \label{eq:hamiltonian_phi}
\end{eqnarray}
where $\tilde{\phi}^{\alpha}_j \equiv \hat{P}^{\alpha} \phi^{\alpha}_j$. 
Here, we reduced the high-energy region ($>\varepsilon_{\mathrm{cut}}$) for approximation.

While we introduced $\hat{P}^{\alpha} \hat{H}^{\alpha} \hat{P}^{\alpha}$ as explained above, we mention here that Eq.~\eqref{eq:hamiltonian} can be generalized as follows:
\begin{equation}
	H_{\alpha' i', \alpha i} \approx
	\langle \lambda^{\alpha'}_{i'} | \hat{P}^{\beta} \hat{H}^{\beta} 
	\hat{P}^{\beta}| \lambda^{{\alpha}}_{i} \rangle,
	\label{eq:hamiltonian-beta}
\end{equation}
where $\beta$ is an arbitrary fragment satisfying $\tilde{\Omega}^{\alpha'} \cap \tilde{\Omega}^{\alpha} \subset \tilde{\Omega}^{\beta}$.
We have found that the choice of $\hat{P}^{\beta} \hat{H}^{\beta} \hat{P}^{\beta}$ has quantitatively no effect on results. 

Combining Eq.~\eqref{eq:hamiltonian} and Eq.~\eqref{eq:hamiltonian_phi}, we can represent the matrix elements of the total Hamiltonian with the fragment orbitals $\{\phi^{\alpha}_i\}$ and the eigenenergies $\{\varepsilon^{\alpha}_i\}$, namely the output of DC-DFT.
Note that the controllable parameters in our eigenstate calculation method are $\varepsilon_{\mathrm{cut}}$, $\lambda_{\mathrm{cut}}$, and the buffer thickness $b'$ of $\tilde{\Omega}^{\alpha}$ ($0 \leq b' \leq b$, see Fig.~\ref{fig:truncation}). 

We need some consideration for the optimum thicknesses $b'$ and $b$. 
The thickness $b$ is dictated by the nearsightedness principle \cite{kohn1996, prodan-kohn} in the conventional DC algorithms. 
On the other hand, the thickness $b'$ is determined by the decay range of the non-local term of the KS Hamiltonian. 
Although large value of $b$ and $b'$ may improve results of the calculation, there is a possibility that the $b' \rightarrow b$ limit degrade the accuracy because of the artificial boundary conditions at $\partial \Omega^{\alpha}$.
Thus it seems that accuracy requirements for our method demand a large $b$ value which can fully contain the nearsightedness range (conventional $b$) plus the Hamiltonian decay range ($b'$). 
However, following results show this is not the case. 
In fact, for the LDC-DFT-based scheme, the $b'=b$ limit leads to the best results in eigenstate calculations (see Sec.~\ref{subsec:parameter}).
This might be attributed to the fact that the uniform kinetic-energy term is the main factor of the non-local part.
Moreover, we can deduce that the periodic boundary conditions at the fragment boundaries $\partial \Omega^{\alpha}$ also improve the accuracy of the eigenstates in condensed matter.
If we utilized other boundary conditions (e.g. insertion of artificial vacuum regions \cite{wang,zhao}), 
the accuracy for the eigenstates might require an optimization of the $b'$ value.

\subsection{Exact exchange potential}

The exact exchange potential or the Hartree-Fock exchange potential $\hat{V}^{\mathrm{HF}}_{\mathrm{x}}$ can be also contained in the total Hamiltonian as a non-local term.
As is well known, this operator is defined as follows,
\begin{eqnarray}
	[\hat{V}^{\mathrm{HF}}_{\mathrm{x}} \psi_i]({\bf r}) 
	&=& -\sum_j^{\mathrm{occ}} \psi_j ({\bf r}) \int d^3 r' 
	\frac{ \psi^{\ast}_j({\bf r}') \psi_i({\bf r}') }{ |{\bf r}-{\bf r}'| } \nonumber \\	
	&=& - \int d^3 r' \frac{ \rho({\bf r},{\bf r}') \psi_i({\bf r}') }{ |{\bf r}-{\bf r}'| },
\end{eqnarray}
where $\{ \psi_i({\bf r}) \}$ are orbitals of the whole system and $\rho({\bf r},{\bf r}')=\sum_i^{\mathrm{occ}} \psi_i ({\bf r}) \psi^{\ast}_i({\bf r}')$ is the density matrix.
The density matrix is exponentially localized with respect to $|{\bf r}-{\bf r}'|$ in many cases  \cite{ismail-beigi}.
Therefore we can take into account only the short-range part of $\hat{V}^{\mathrm{HF}}_{\mathrm{x}}$ and adopt the DC scheme in such cases.
It is expected that our method allows high-speed eigenstate calculations within hybrid functionals.
In a naive implementation of the exact exchange, the computational cost is proportional to the fourth power of the system size.
On the other hand, in the present method, the computational cost of the Hamiltonian construction is negligible and the cost of the diagonalization is proportional to the third power of the small matrix dimension as explained in Sec.~\ref{subsec:strategy}.

In Sec.~\ref{subsec:time}, we demonstrate the validity of our scheme to construct the Hamiltonian with the exact exchange potential. 
Namely, after doing the conventional calculation including the exact exchange in each fragment independently,
we construct the basis set $\{|\lambda^{\alpha}_i\rangle\}$ and the Hamiltonian matrix $\{ H_{\alpha' i', \alpha i} \}$ from the orbitals derived by these calculations. 
We shall see that the energy eigenvalues are accurately reproduced (see Sec.~\ref{subsec:time}).
More thorough divide-and-conquer calculations with the exact exchange will be presented elsewhere \cite{future}.

\subsection{Computational details}

We have implemented LDC-DFT and our method in the xTAPP code \cite{xtapp}.
The LDC-DFT code is parallelized using the message passing interface (MPI) library.
To perform our procedure, we utilized the grid points of the fast Fourier transformations (FFT) for the projection and the inner-product operations of the fragment orbitals.
The process to generate the basis functions is completely parallelized, while the Hamiltonian matrix operation needs MPI communications for calculating the off-diagonal blocks.
The matrix elements of the Hamiltonian are gathered in the root process, and then the matrix is diagonalized by a serial computation.
In the eigenstate calculations, we shifted the energy origin as
$\varepsilon^{\alpha}_i \longrightarrow \varepsilon^{\alpha}_i - \varepsilon_{\mathrm{cut}}$
to suppress a contamination by $0$ matrix elements.
We can obtain final results through reshifting,
$\varepsilon_i \longrightarrow \varepsilon_i + \varepsilon_{\mathrm{cut}}$ ,
where $\{ \varepsilon_i \}$ are eigenenergies of the total Hamiltonian.

In the following calculations, we used the plane-wave basis (for calculations in each fragment), the norm-conserving pseudopotentials, and PBE exchange-correlation functional \cite{pbe} except some calculations with the exact exchange, in which PBE0 hybrid functional \cite{pbe0} was used.
All the calculations were carried out in the paramagnetic case.
For comparison we also performed the conventional DFT calculations, where we sampled the Brillouin zone at the $\Gamma$ point (${\bf k}=0$ point).

\section{Results and Discussion \label{sec:results}}

\subsection{Parameter dependence \label{subsec:parameter}}

\begin{figure}
	\includegraphics[keepaspectratio, scale=0.6]{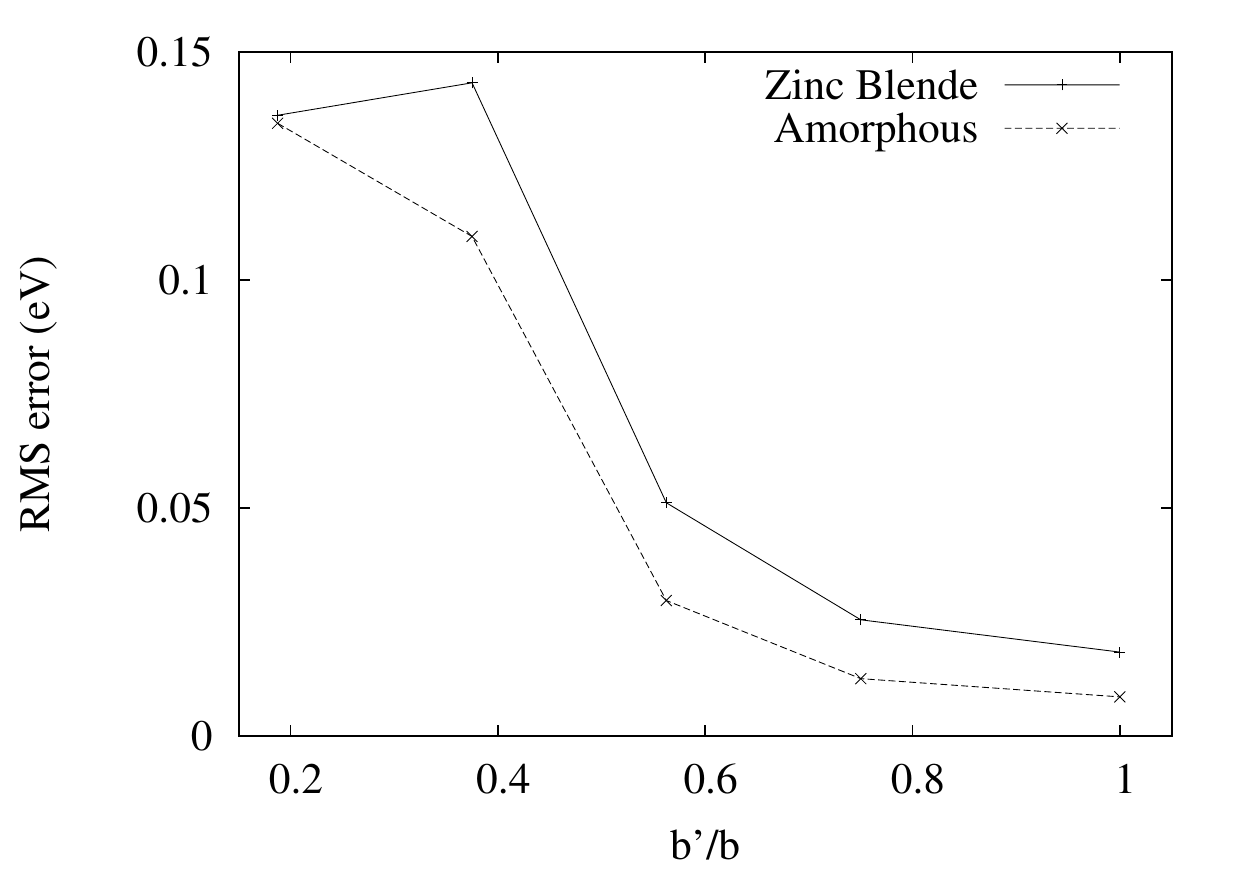}
	\caption{\label{fig:b} 
		RMS error of the occupied eigenenergies as a function of a ratio $b'/b$
		 for SiC systems with zinc-blende structure and amorphous structure, 
		 where each system contains 512 atoms in a cubic $4 \times 4 \times 4$ supercell.
		 The other parameters are fixed as $\varepsilon_{\mathrm{cut}}-\mu=10.88$ eV (0.4 Hartree)
		 and $\lambda_{\mathrm{cut}}=10^{-3}$. 
	}
\end {figure}

\begin{figure}
	\includegraphics[keepaspectratio, scale=0.6]{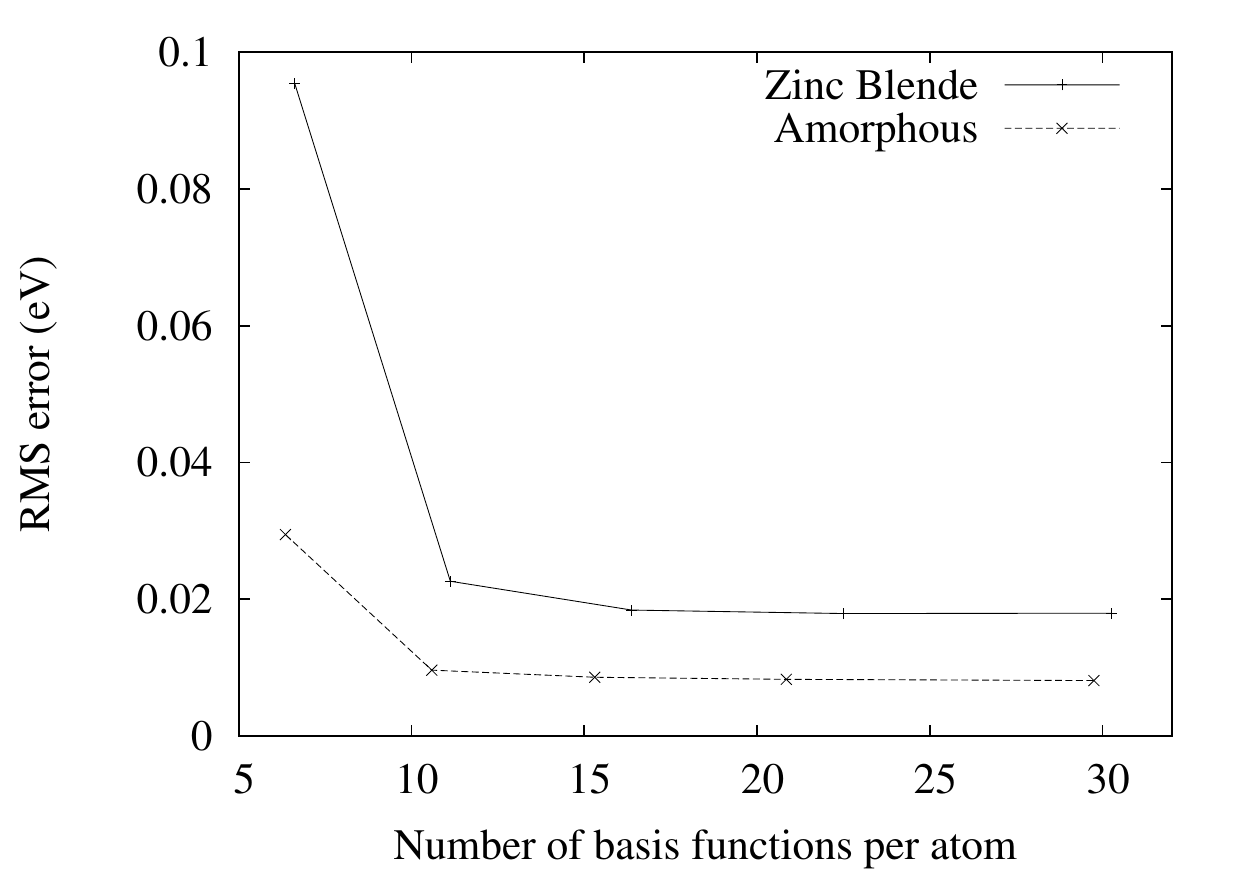}
	\caption{\label{fig:lcut} 
		RMS errors plotted against a number of the basis functions per atom	
		for the same systems as Fig.~\ref{fig:b},
		where $b'=b$ and $\varepsilon_{\mathrm{cut}}-\mu=10.88$ eV (0.4 Hartree).
		These points correspond to $\lambda_{\mathrm{cut}}=10^{-1}$, $10^{-2}$, $10^{-3}$,
		$10^{-4}$, and $10^{-5}$ respectively from left to right. 
	}
\end {figure}

\begin{figure}
	\includegraphics[keepaspectratio, scale=0.6]{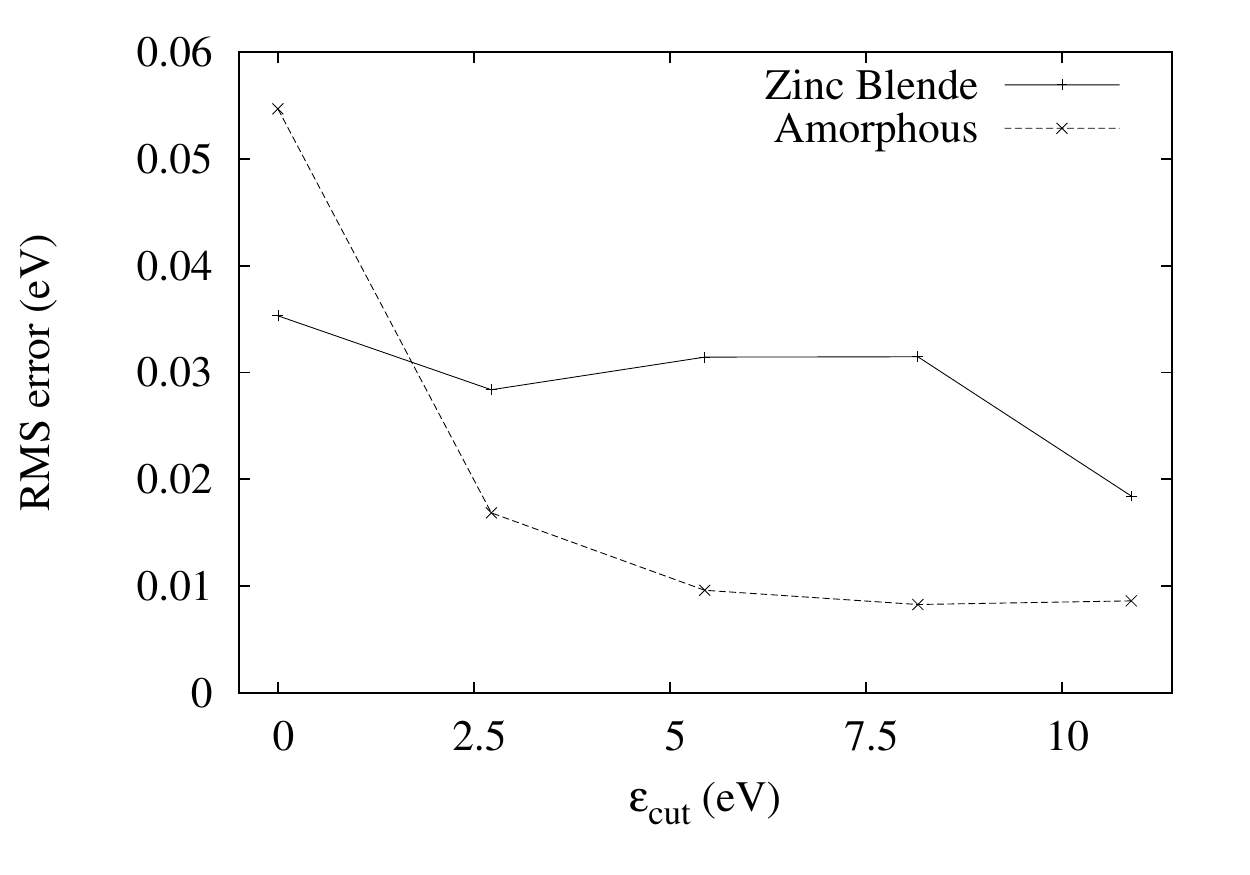}
	\caption{\label{fig:ecut} 
		The $\varepsilon_{\mathrm{cut}}$ dependence of the RMS errors 
		for the same systems as Fig.~\ref{fig:b},
		where the energy origin is fixed at the Fermi energy $\mu$.
		The other parameters are fixed as $b'=b$ and $\lambda_{\mathrm{cut}}=10^{-3}$.
	}
\end {figure}

We examine a parameter dependence for the resultant eigenenergies of the present method with respect to $b'$, $\lambda_{\mathrm{cut}}$ and $\varepsilon_{\mathrm{cut}}$. 
Trial systems are a SiC 512-atom systems with zinc blende structure and an amorphous structure
 ($4 \times 4 \times 4$ supercells). 
We divide the systems into 64 fragments (with $4 \times 4 \times 4$ configurations), respectively,  whose side lengths of each cubic core domain and the buffer thickness $b$ are fixed as $a=4.39$ {\AA}  (experimental lattice constant).
The plane-wave cutoff is 30 Ry and the number of the FFT mesh points is equal to $16 \times 16 \times 16$ in each core domain.

Figure~\ref{fig:b} shows the relation between the thicknesses $b'$ and the root mean square (RMS) errors of the occupied eigenenergies with respect to the conventional results. 
Here, RMS error of $n$ eigenenergies is defined as follows:
\begin{equation}
\mathrm{RMS\,\, error} = \sqrt{ \frac{1}{n} \sum_i^n (\varepsilon_i - \varepsilon^0_i)^2},
\end{equation}
where $\varepsilon_i$ and $\varepsilon_i^0$ are the orbital energies for the whole system obtained by our method and the conventional DFT method, respectively.
This figure indicates that $b'=b$ leads to the best results for not only the zinc-blende structure but also the amorphous structure, although there was the concern of boundary effects in the latter case.
With these results, we put $b'=b$ in all the following calculations.

In order to examine the $\lambda_{\mathrm{cut}}$ dependence, we change $\lambda_{\mathrm{cut}}$ value from $10^{-1}$ to $10^{-5}$.
Figure~\ref{fig:lcut} is the RMS errors plotted against the number of the basis functions ($\sum_{\alpha} M_{\alpha}$) per atom with the same conditions, where $\varepsilon_{\mathrm{cut}}-\mu=10.88$ eV (0.4 Hartree).
The RMS errors are saturated at nearly 15 basis functions per atom when $\lambda_{\mathrm{cut}}=10^{-3}$.

Figure~\ref{fig:ecut} illustrates the $\varepsilon_{\mathrm{cut}}$ dependence of the RMS errors with the same conditions, where the energy origin is fixed at the Fermi energy $\mu$ (= the valence band maximum in a gapped system).
In the amorphous structure, the eigenenergies of the occupied states are accurately evaluated with a large $\varepsilon_{\mathrm{cut}}$ value.
On the other hand, the error for the zinc-blende structure indicates weak dependence on $\varepsilon_{\mathrm{cut}}$ probably due to the band gap.
In a gapped state, we can deduce that it is sufficient to utilize the occupied fragment orbitals for representing the occupied states of the total system.

In summary, the accuracy for the eigenstate requires the parameter values for $b'=b$, $\lambda_{\mathrm{cut}} \approx 10^{-3}$, and a sufficiently large $\varepsilon_{\mathrm{cut}}$ value compared to the desired energy range.
These conditions provide the accuracy comparable with FMO-LCMO \cite{tsuneyuki, kobori}.

\subsection{Phosphorous-doped Silicon}

\begin{figure*}
	\subfloat[]{
		\includegraphics[keepaspectratio, scale=0.4]{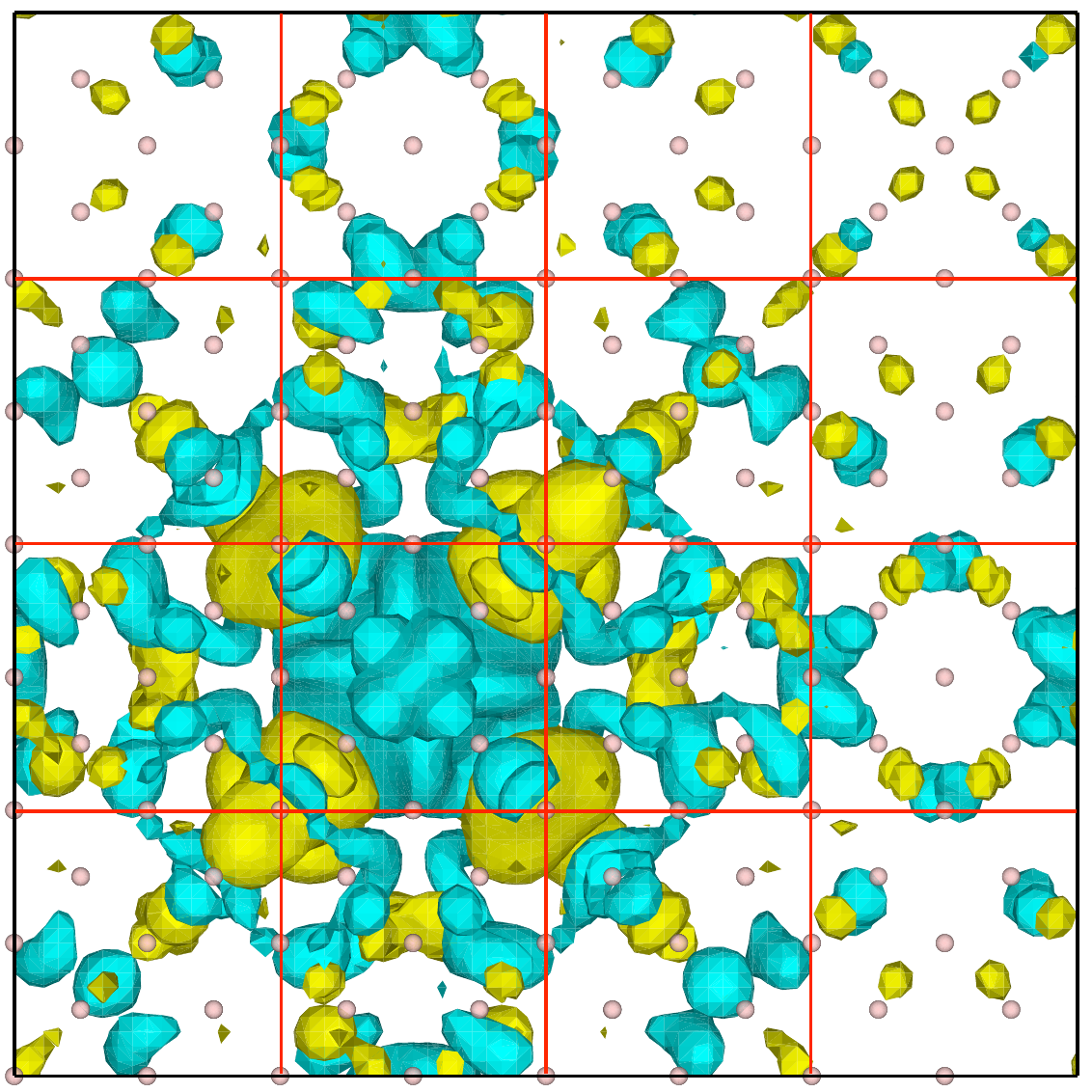}
	}
	\subfloat[]{
		\includegraphics[keepaspectratio, scale=0.4]{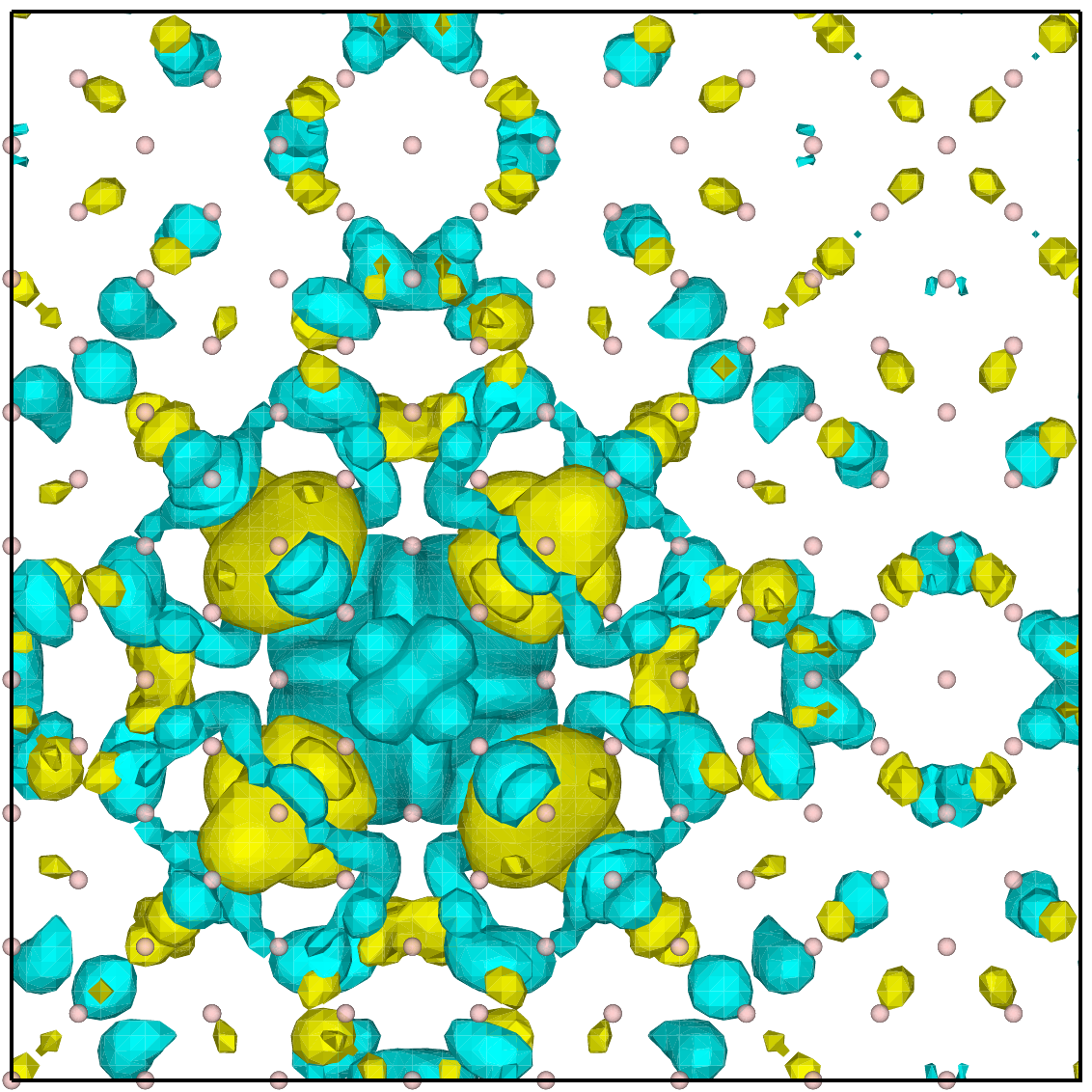}
	}
	\caption{
		(a) The DC-LCFO wave function of the donor state for the P-doped Si 
		(using VESTA \cite{momma}).
		The system contains 511 Si atoms and 1 P atom in a cubic $4 \times 4 \times 4$ supercell.
		The red lines indicate the core domains $\{ \Omega_{0}^{\alpha} \}$.
		(b) The conventional result for comparison.
	}
	\label{fig:wf}
\end{figure*}

We demonstrate that the present method can represent a defect state with a spatially extended wave functions with satisfactory accuracy.
To this end, we perform a calculation for a P-doped Si crystal that contains 512 atoms (one P atom included) and divide the system into
64 fragments with the $4 \times 4 \times 4$ configuration, where the side lengths of each core domain and the buffer thickness are fixed to $a=5.43$ {\AA} (experimental lattice constant). 
The controllable parameters of the eigenstate calculation are set as $b'=b$, $\lambda_{\mathrm{cut}}=10^{-3}$, and $\varepsilon_{\mathrm{cut}}-\mu=8.163$ eV ($0.3$ Hartree). 
The plane-wave cutoff is 30 Ry and the number of the FFT mesh points is equal to $18 \times 18 \times 18$ in each core domain.

The RMS error and the maximum absolute error (MAE) for the occupied eigenenergies ($\{ \varepsilon_i \}^{1025}_{i=1}$) are 0.013 eV and 0.084 eV, respectively.
Here, $\varepsilon_i$ is the eigenenergy of the total system and the highest occupied state is the half filled donor state ($i=1025$).
The absolute error for the donor state eigenenergy is 0.004 eV. 
The RMS error and the MAE for the unoccupied eigenenergies ($\{ \varepsilon_i \}^{1100}_{i=1026}$) are 
0.061 eV and 0.133 eV, respectively.
The number of the basis functions, or the dimension of the Hamiltonian matrix, is $9046$.
Figure~\ref{fig:wf} shows the wave function of the donor state calculated with our method and the conventional method for comparison.
From this result it can be seen that our scheme can properly represent the wave function extended over fragments.

\subsection{InGaN/GaN superlattice}

\begin{figure*}
	\includegraphics[keepaspectratio, scale=0.8]{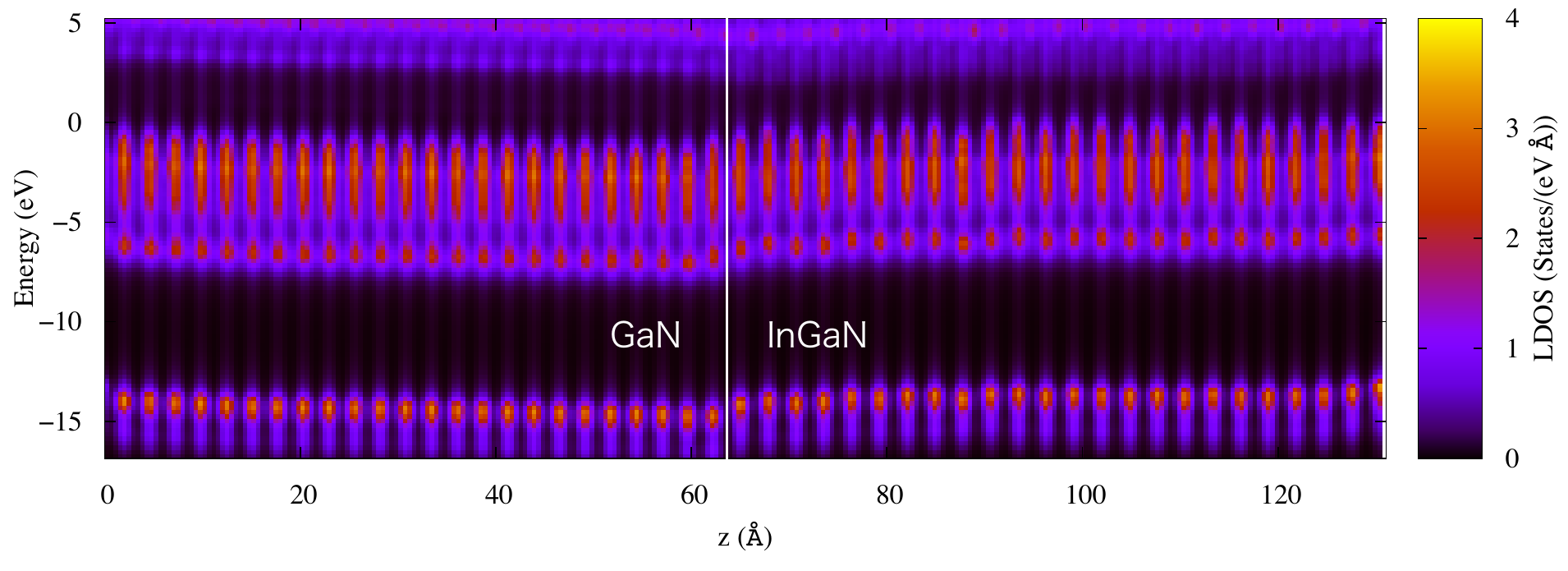}
	\caption{
		The local density of states (LDOS) of the InGaN/GaN superlattice system,
		which is composed of 24 InGaN layers and 24 GaN layers, 
		where each layer contains 16 atoms.
		The system is divided into 12 fragments along the z-direction.
	}
	\label{fig:ldos}
\end {figure*}

We applied our method to a superlattice system with the polar interface.
InGaN/GaN heterostructures with [0001] epitaxial alignments have such interface which induces band bending \cite{dong}.
We show results of the eigenstates calculation for (In${}_{0.5}$Ga${}_{0.5}$N)${}_{24}$/(GaN)${}_{24}$ superlattice which contains 768 atoms in a rectangular cell of side lengths, $12.89 \times 5.58 \times 131.06$ (in \AA).
The system is divided into 12 fragments with 1D ($1 \times 1 \times 12$) configurations. 
The buffer thickness $b$ is equal to the length of each core domain along the $z$ direction, while the $x$-$y$ plane is not divided so $b=0$ in the $x$-$y$ plane.
The parameters for the eigenstate calculation are fixed as $b'=b$, $\lambda_{\mathrm{cut}}=10^{-3}$, and $\varepsilon_{\mathrm{cut}}-\mu=8.163$ eV ($0.3$ Hartree).
The plane-wave cutoff is 50 Ry and the number of the FFT mesh points is equal to $60 \times 24 \times 48$ in each core domain.

The RMS error and the MAE for the occupied eigenenergies ($\{ \varepsilon_i \}^{1536}_{i=1}$) are 0.031 eV and 0.186 eV, respectively, 
while the RMS error and the MAE for the unoccupied eigenenergies ($\{ \varepsilon_i \}^{1700}_{i=1537}$) are 0.027 eV and 0.084 eV, respectively.
The dimension of the total Hamiltonian matrix in this calculation is 3301 (i.e. the number of the basis functions per atom is nearly equal to 4.3).
This small number of the basis functions gives sufficient accuracy because of the 1D configuration of the fragments where each fragment has only two boundary regions in contrast to 26 in case of 3D configuration.
Figure~\ref{fig:ldos} shows the local density of states (LDOS) indicating the band-bending structure.

\subsection{Benchmark tests for the Hamiltonian diagonalization \label{subsec:time}}

\begin{figure}
	\includegraphics[keepaspectratio, scale=0.6]{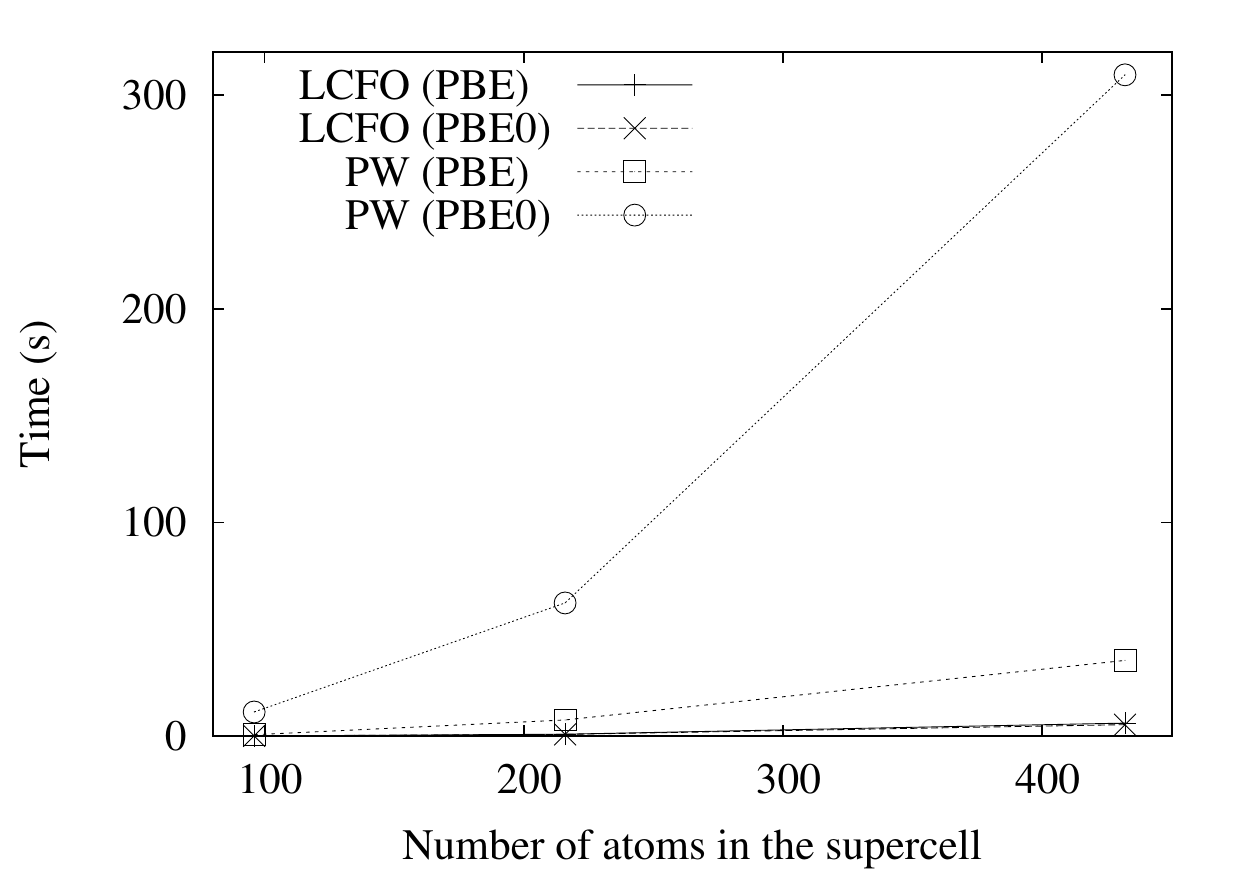}
	\caption{\label{fig:time} 
		The computational cost of the Hamiltonian diagonalization
		(per iteration for the conventional cases) 
		with different supercells, where the exchange-correlation functional is PBE or PBE0.
		LCFO indicates our method
		, while PW corresponds to the conventional plane-wave calculation.
		The computation is done on a 24-cores 2.5 GHz Intel Xeon cluster with 144 nodes.
	}
\end {figure}

Finally we performed benchmark tests of the DC-LCFO method using Si 96-atom ($12 \times 1 \times 1$), 216-atom ($27 \times 1 \times 1$), and 432-atom ($54 \times 1 \times 1$) supercells.
The systems are divided into 12, 27, and 54 fragments along the $x$ direction, respectively.
We set $b=a$ (lattice constant) for each fragment so that it is equal to $3 \times 1 \times 1$ supercell.
The basis set $\{|\lambda^{\alpha}_i\rangle\}$ and the Hamiltonian matrix $\{ H_{\alpha' i', \alpha i} \}$
are derived from copied wave functions, which are obtained by the conventional calculation in the $3 \times 1 \times 1$ cell, where we used PBE and PBE0 functionals.
We fixed the parameters as $b'=b$, $\lambda_{\mathrm{cut}}=10^{-3}$, and $N_{\alpha}=56$ (= occupied + 8 orbitals, instead of $\varepsilon_{\mathrm{cut}}$).
For the PBE0 calculation, we used the Coulomb potential cutoff $R_{\mathrm{c}}$ \cite{spencer, matsushita} for truncating $\hat{V}^{\mathrm{HF}}_{\mathrm{x}}$ and set $R_{\mathrm{c}}=8$ a.u. $<b'$ as a simple implementation.

Figure~\ref{fig:time} shows the computational cost for the construction and diagonalization of the Hamiltonian (per iteration for the conventional cases).
This result suggests that our method enables a high-speed computation of the eigenstate.
As reference, we describe the RMS error of the occupied eigenenergies and the error of the band gap
for the PBE0 hybrid functional in $54 \times 1 \times 1$ cell;
0.034 eV and 0.029 eV, respectively.

\section{Conclusion \label{sec:conclusion}}

We have developed a method, named DC-LCFO, to compute the orbital wave functions and the corresponding orbital energies of general huge systems based on DC-DFT.
The method utilizes the output derived by DC-DFT for constructing the total Hamiltonian matrix without redundant computations. 
Furthermore, this method can dramatically reduce the matrix size of the total Hamiltonian.
Thus its computational cost is much lower than the conventional calculations.
We have applied the method to P-doped Si and InGaN/GaN superlattice systems to demonstrate that it reproduces a structure of the wave functions spread over the total system with practical accuracy.

DC-LCFO is a powerful tool for studying the nature of electronic states, the mechanism of a chemical reaction for example, of large systems.
In principle, our scheme can be applied to the Hamiltonian including short-range exchange-correlation terms such as the GW self-energy operator \cite{hedin,hedin1969,hybertsen}.

\begin{acknowledgments}
This work was supported in part by JSPS KAKENHI Grant Number 26286085, and MEXT as a social and scientific priority issue (Creation of new functional devices and high-performance
materials to support next-generation industries; CDMSI) to be tackled by using post-K computer.
S. Y. was supported by the advanced leading graduate course for photon science (ALPS).
The authors thank the Supercomputer Center, the Institute for Solid State Physics, the University of Tokyo for the use of the facilities.
\end{acknowledgments}

\appendix

\section{Comparison with FMO-LCMO \label{appendix}}

\begin{figure}
	\includegraphics[keepaspectratio, scale=0.75]{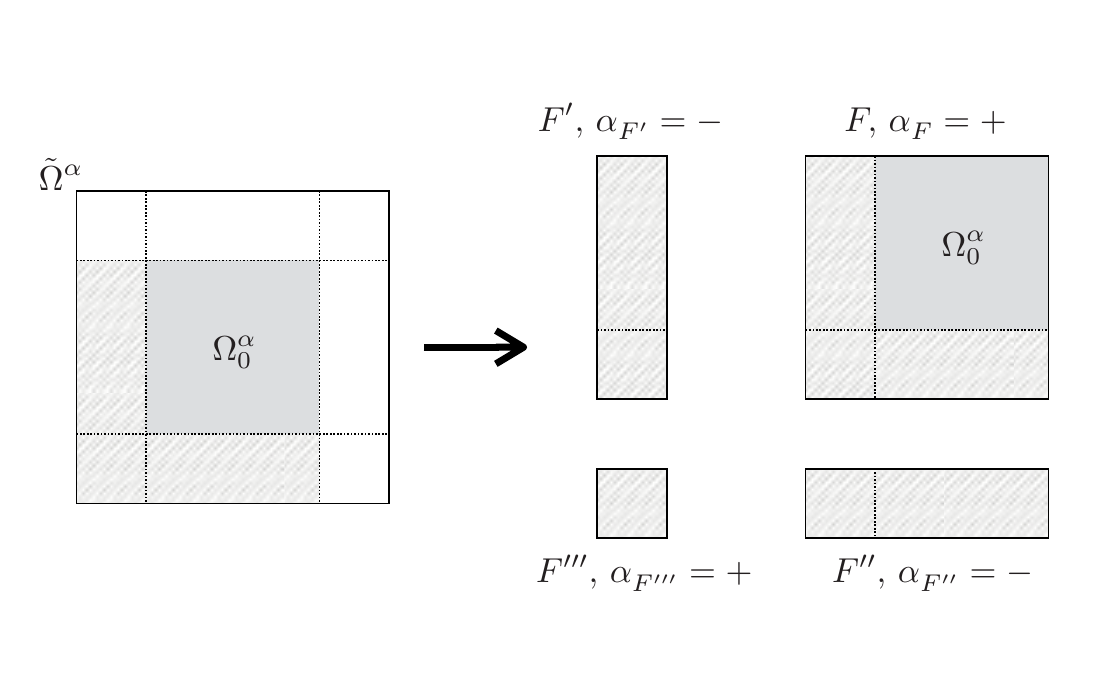}
	\caption{\label{fig:ls3df} 
		LS3DF-like decomposition of the fragment Hamiltonian $\hat{H}^{\alpha}$.
	}
\end {figure}

In Sec.~\ref{subsec:hamiltonian}, we formulated the approximate expression for the Hamiltonian matrix, starting from the consideration of the Hamiltonian operation on the basis functions.
In FMO-LCMO, on the other hand, the approximate Hamiltonian is represented as a decomposed operator which consists of the fragment Hamiltonian operators.
Let us discuss the relation between these two formulations.

In FMO-LCMO, the total Hamiltonian is first decomposed into the "fragment monomer" and "fragment dimer" (the union of two monomers) terms as follows \cite{tsuneyuki,kobori}:
\begin{equation}
	\hat{H}=\sum_I \hat{H}_I + \sum_{I>J} \left (\hat{H}_{IJ} - \hat{H}_I - \hat{H}_J \right),
	\label{eq:fmo-lcmo}
\end{equation}
where we omit the higher terms such as the "fragment trimer" (FMO2 level). 
$I$ and $J$ are indices of the fragment monomers and $\hat{H}_I$ denotes the Hamiltonian for the fragment monomer $I$. 
$\hat{H}_{IJ}$ represents the Hamiltonian for the fragment dimer $IJ$.
The second term represents the non-local effects across the monomers, while the first term indicates the monomer effects.     

One can construct a representation analogous to Eq.~\eqref{eq:fmo-lcmo} for the Hamiltonian operator so that its matrix elements agree with Eq.~\eqref{eq:hamiltonian-beta}.
This representation is formulated by the summation of the fragment Hamiltonians in a similar way to the sum formula of electron density in a DC approach called LS3DF \cite{wang,zhao}.
For simplicity, we consider only a case of 2D systems. 
Here, each $\tilde{\Omega}^{\alpha}$ is divided into 4 small fragments and they are assigned new indices $F$.
For each fragment $F$, we assign a sign factor $\alpha^{}_F=\pm$ depending on the layout of the fragment (Fig.~\ref{fig:ls3df}).
The approximated Hamiltonian operator of the whole system can be expressed as,
\begin{equation}
	\hat{H}_{\mathrm{approx.}}=\sum_F \alpha^{}_F \hat{H}^{}_F,
\end{equation}
where $\hat{H}_F=\hat{P}^F \hat{H}^{\alpha} \hat{P}^F$ is a projected Hamiltonian for the small fragment $F$.

With the above $ \hat{H}_{\mathrm{approx.}}$, the matrix elements are completely identical to $ \langle \lambda^{\alpha'}_{i'} | \hat{P}^{\beta} \hat{H}^{\beta} \hat{P}^{\beta}| \lambda^{{\alpha}}_{i} \rangle $ with a proper configuration of the fragment $\beta$. 
Specifically, we put $\beta = \alpha$ ($\alpha'$) if the fragment $\alpha$ is located at the north/north-east/east (south/south-west/west) of the fragment $\alpha'$ in 2D systems, while $\beta$ indicates the other fragment overlapping with both $\alpha$ and $\alpha'$ if the fragment $\alpha$ is located at the north-west/south-east of the fragment $\alpha'$ as Fig.~\ref{fig:beta-other}.

%


\end{document}